\documentclass{elsart}

\usepackage{amssymb,graphicx}

\def\be{\begin{equation}}
\def\ee{\end{equation}}

\begin{document}

\begin{frontmatter}



\title{Routes to chaos, universality and glass formation}


\author{Fulvio Baldovin}

\address{
Dipartimento di Fisica and
Sezione INFN, Universit\`a di Padova,\\
\it Via Marzolo 8, I-35131 Padova, Italy
}

\ead{baldovin@pd.infn.it}

\begin{abstract}
We review recent results obtained for the dynamics of
incipient chaos. These results suggest a common picture underlying the
three universal routes to chaos displayed by the prototypical
logistic and circle maps. Namely, the period doubling,
intermittency, and quasiperiodicity routes. In these situations
the dynamical behavior is exactly describable through infinite
families of Tsallis' $q$-exponential functions.
Furthermore, the addition of a noise perturbation to the dynamics at
the onset of chaos of the logistic map allows to establish parallels
with the behavior of supercooled liquids close to glass formation.
Specifically, the occurrence of two-step relaxation, aging with its
characteristic scaling property, and subdiffusion and arrest is
corroborated for such a system.
\end{abstract}

\end{frontmatter}

% main text
\section{Introduction}
In recent years there has been a renewed interest in the analysis of
the dynamical behavior at critical attractors in low dimensional maps. In
such situations, criticality is triggered by the vanishing of the
Lyapunov coefficient (coefficients in dimension higher than one) so
that ordinary chaos paradigms are not applicable.  Part of this
interest is because Tsallis' functional forms (the
so-called $q$-exponential and $q$-logarithm) were found to mimic
the role that the ordinary exponential and
logarithm have in the usual theory of chaos (although, as we will see
below, in a more complex and sophisticated sense).  
In fact, exact renormalization group (RG) solutions for the
sensitivity to initial conditions at these critical attractors are
found to assume the structure of (infinite families of)
$q$-exponential functions.  
This paper attempts to rationalize in a single review some of the
most relevant new results that stemmed from the application of
exact RG approaches to critical attractors of low dimensional maps
\cite{robledo_1,baldovin_1,baldovin_2,baldovin_3,mayoral_1,saldana_1,robledo_2,baldovin_4}.
This research work moved from the understanding that RG techniques
developed in the '80s for describing the static properties of the maps
could indeed be extended in order to solve in all details the dynamical
behavior at critical attractors. Then, contact with Tsallis' and Mori's
formalisms were established {\it a posteriori}. The final result is 
the identification of a universal,
intricate, dynamical mechanism common to the three paradigmatic
routes to chaos of period doubling, intermittency and
quasiperiodicity. 

Another important result conveyed by the RG approach at the chaos
threshold of the logistic map is the 
determination of the exponent characterizing the decay of correlations
in presence of an external noise source \cite{robledo_2,baldovin_4}. 
By realizing that such an
iteration equation is a discrete version of Langevin's equation and
that the underlying scenario is that of an ergodic to non-ergodic
transition, one can in fact produce a ``translation dictionary''
between the dynamical behavior of the logistic map with additive noise
and the phenomenology observed for a glass former. 
Translated words include Adam-Gibbs formula, 
time translation invariance and $\alpha$-relaxation,
aging, subdiffusion and arrest.

There is a further interesting mechanism through which aging and weak
ergodicity breaking can be obtained in low dimensional maps. This
mechanism is related to the dynamical behavior close to the origin 
of a Manneville-like map (see below) and can be statistically
described in terms of the continuous time random walk formalism. Such
a discussion is however out of the scope of the present paper. The
interested Reader is referred to \cite{barkai} for an appropriate
account.

The paper is divided in two main sections. Section \ref{routes} is
devoted to the dynamical picture underlying the above mentioned routes to
chaos. In Section \ref{glassy} we describe the
analogies between the behavior of the logistic map with additive
noise and that of a glass former. Finally, some remarks
are added.

\section{Routes to chaos: a common dynamical picture}
\label{routes}
\begin{figure}
\begin{center}
\includegraphics[width=0.50\columnwidth]{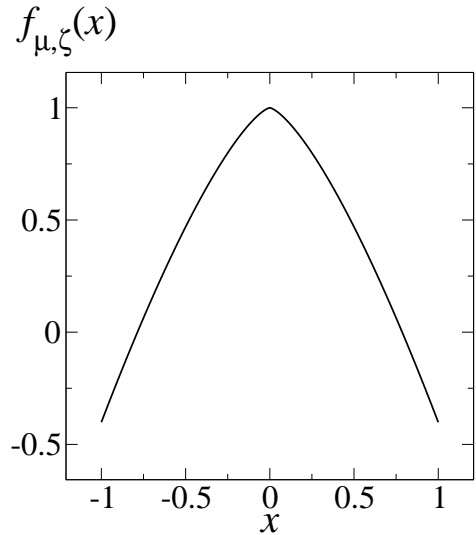}
\caption{
Unimodal map.
}
\label{fig_unimodal}
\end{center}
\end{figure}

In the following, we will be mainly considering unimodal maps, i.e.,
discrete-time iterated maps governed by the equation
\begin{equation}
{x_{t+1}} =  f_{\mu,\zeta}(x_t)\equiv { 1-\mu
|x_t|^\zeta},\qquad t=0,1,\ldots,\qquad x\in[-1,1],
\label{eq_unimodal}
\end{equation}
where $\mu\in[0,2]$ is the control
parameter and $\zeta>1$ the non linearity order of the
maximum at $x=0$ (Fig. \ref{fig_unimodal}). 
$\zeta=2$ corresponds to the celebrated logistic map
\cite{schuster,beck}. 
We define the sensitivity to initial conditions as
\be
{\xi(x_0, t)}\equiv\lim_{|\Delta x_0|\to0}\frac{|{ \Delta
    x_t}|}{|{ \Delta x_0}|},
\ee 
where we denote explicitly the dependence on the initial condition
$x_0$ 
in order to address cases in
which the Lyapunov coefficient is zero.  
Typically (for all $\mu$ except a set of zero
Lebesgue measure), $\xi$ is independent from
$x_0$ and exponential for $t\gg1$, ${ \xi(t)\sim\exp\left(\lambda t\right)}$,
where $\lambda\in\mathbb R$ is the Lyapunov coefficient \cite{schuster,beck}. 

\begin{figure}
\begin{center}
\includegraphics[width=0.48\columnwidth]{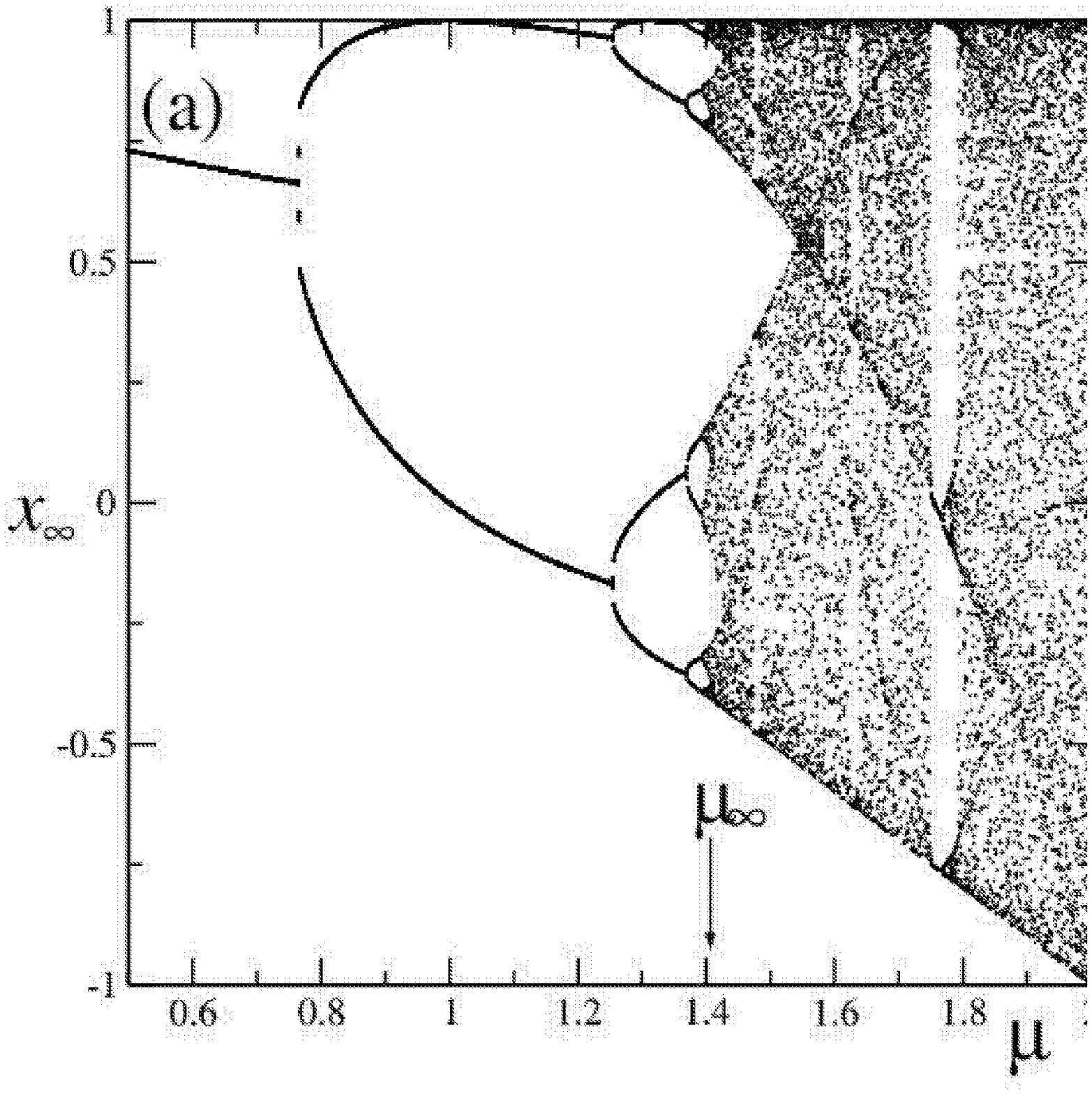}
\includegraphics[width=0.48\columnwidth]{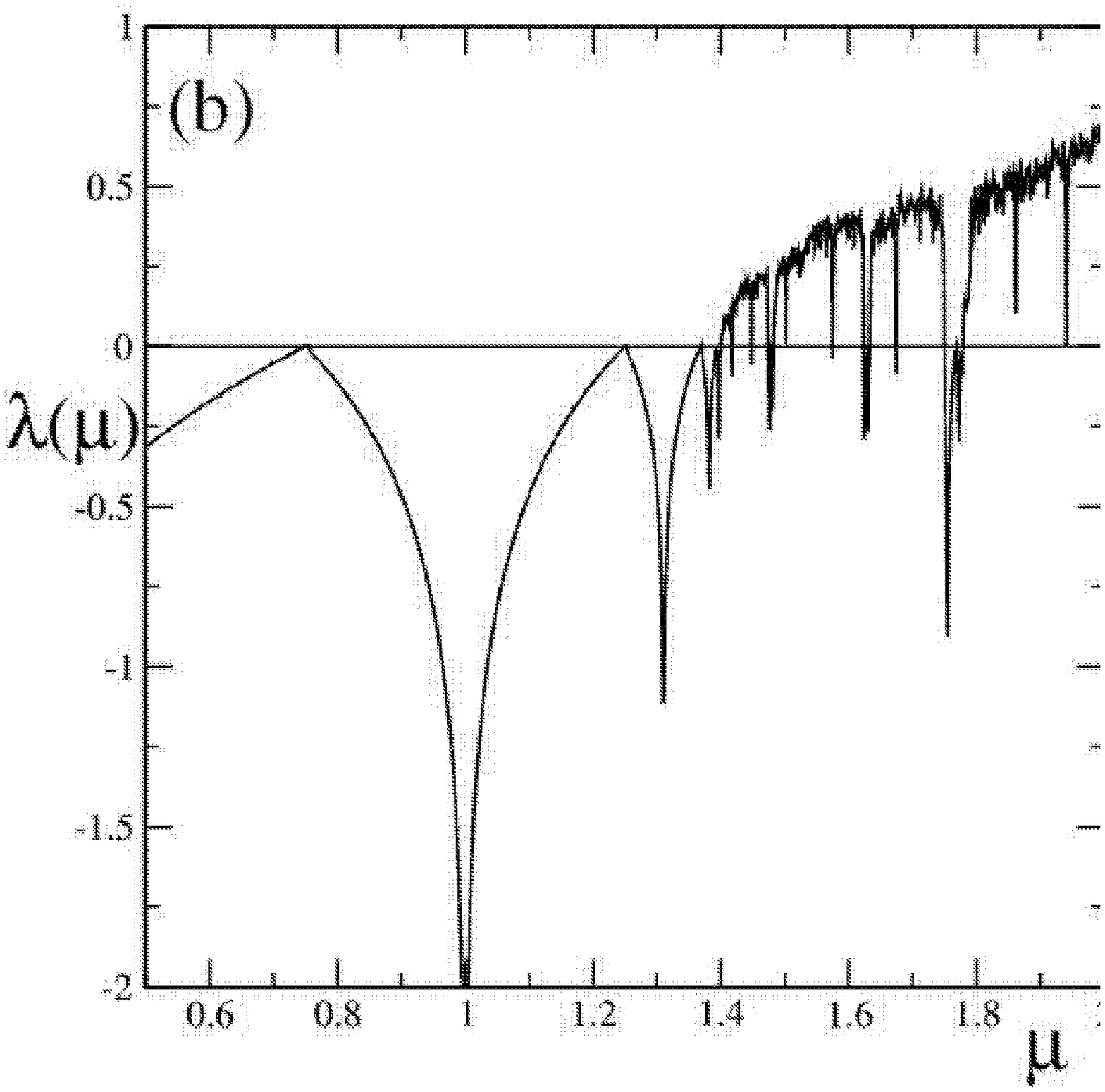}
\caption{
Attractor (a) and Lyapunov structure (b) of the logistic map ($\zeta=2$).
}
\label{fig_attractor}
\end{center}
\end{figure}

For $\mu$ smaller than a critical value
$\mu_\infty(\zeta)$, $\lambda$ is negative except for an infinite but numerable
number of pitchfork bifurcations points in which $\lambda=0$
(see Fig. \ref{fig_attractor}). 
Within this region, the attractor is periodic of order $2^r$
($r=0,1,\ldots$) with $r$
that increases by one unit at each pitchfork bifurcation and that becomes
infinite at $\mu_\infty$.   
Inside each $\mu$-interval in which the period is $2^r$, there is a special
orbit that passes through $x=0$ and that has $\lambda\to-\infty$
(super-stable cycle \cite{schuster}). 
Important quantities for the
dynamics at the onset of chaos are the diameters of the bifurcation
forks $d_{r,s}$ ($s=0,1,\ldots,2^r-1$), which are defined as the
distances of two nearest neighbors attractor points in a super-stable
cycle. As we will see, in the limit $r\to\infty$
these diameters contain all the information about
the multifractal attractor at the onset of chaos.   
The chaos threshold $\mu_\infty$ is the accumulation point
of both pitchfork bifurcations and super-stable cycles.

For $\mu>\mu_\infty$, $\lambda$ is positive  except
in an numerable number of intervals which are again characterized by a
periodic attractor. 
The left extreme of such intervals is determined by a tangent
bifurcation where $\lambda$ vanishes.
Inside each interval, a new pitchfork bifurcation
cascades produces a (higher order) chaos threshold before reaching the
right extreme of the interval, so that the final result is an
intricate and fascinating self-similar structure.   

The study of unimodal maps marked an era in chaos theory. For example, the 
pitchfork cascade has been identified to occur in chemical
reactions, optically bistable systems, electrical RCL oscillators,
sound waves in water, B\'enard convection fluid experiments, \ldots
\cite{schuster}.  
Also, the tangent bifurcation with
the addition of a reinjection mechanism is the paradigm for the
phenomenon of (type I) intermittency and $1/f$-noise 
observed, e.g., in nonlinear RCL oscillators and in B\'enard
convection \cite{schuster}.  
The elegance and richness of this self-similar structure stimulated a
number of theoretical results starting from the late $70$'s.
Perhaps the milestone was the RG approach developed by Feigenbaum
\cite{feigenbaum} and, independently, by Coullet and Tresser \cite{coullet} for
the pitchfork bifurcations cascade. This work was extended by
the group of Politi \cite{politi} and that of Mori
\cite{mori_1} that focused on the properties of the fluctuation
spectrum of generalized algebraic Lyapunov coefficients. 
An exact solution for the tangent bifurcation dynamics was found
by Hu and Rudnick \cite{hu}. 
As we show below, such solution can be extended also to
the pitchfork bifurcations. 
Another important contribution by Mori and
colleagues was the introduction of the idea of  
{\it dynamical transitions} \cite{mori_2}, 
a formalism closely related to the probabilistic large
deviation theory \cite{oono}, that is analogous to the description of
first-order thermal transitions.  
More recently, the group of Tsallis \cite{tsallis} pointed out that the
$q$-exponentials derived within the so-called nonextensive statistical
mechanics formalism appropriately describe the envelope of the
fluctuating sensitivity to initial conditions in which the ordinary
Lyapunov coefficient vanishes. 
The work by Robledo and colleagues 
\cite{robledo_1,baldovin_1,baldovin_2,baldovin_3,mayoral_1,saldana_1} 
that we are reviewing in this
section put together all previous approaches in a single 
perspective, by deriving exact results supported by the RG functional
composition.   

\subsection{Pitchfork and tangent bifurcations}
\begin{figure}
\begin{center}
\includegraphics[width=0.60\columnwidth]{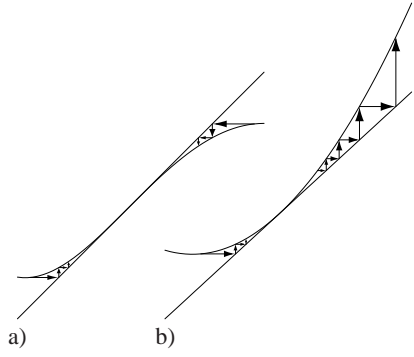}
\caption{
Pitchfork (a) and tangent (b) bifurcations.
}
\label{fig_bifurcations}
\end{center}
\end{figure}

Let us start by understanding what happens at  pitchfork and
tangent bifurcations. Reading such a transition along the $\mu$-axis
of Fig. \ref{fig_attractor}a, the transition 
is said of order $n$ if in the left (pitchfork) or in the right
(tangent) neighborhood of the transition
point the attractor is periodic of order $n$.
If we consider the $n$-composed map $f_{\mu,\zeta}^{(n)}$ and shift
the coordinate $x$ to one of the of the $n$ bifurcation points\footnote{
With an abuse of notations we still indicate the shifted coordinate
$x^\prime$ as $x^\prime\equiv x$.
}, we have
that in the neighborhood of the chosen point $f_{\mu,\zeta}^{(n)}$ has
the universal expansion 
\begin{equation}
f_{\mu,\zeta}^{(n)}(x)=x+(-1)^z u\;{\rm sgn}(x^z)|x|^{z}+o(|x|^{z}),
\label{eq_manneville}
\end{equation}
where $u>0$ is the leading expansion coefficient and ${\rm sgn}$ is
the sign function (here and below repeated use of the sign function
is made in order to unify the treatment of pitchfork and tangent
bifurcations). 
For a pitchfork bifurcation $z=3$, whereas for a tangent bifurcation
$z=2$ (see Fig. \ref{fig_bifurcations}). 
The expansion is independent of the non-linearity parameter
$\zeta$. 

With $z=2$, by neglecting the $o(|x|^z)$ term, Eq. (\ref{eq_manneville})
recovers the Manneville map \cite{schuster}. 
In such a case, at the left side of the bifurcation
point trajectories are converging to the bifurcation point, whereas at the
right side they are expelled from the bifurcation point
(Fig. \ref{fig_bifurcations}b). 
As we will see below, in the converging side the dynamics is regular,
being characterized by a weak (power law in place of exponential) 
{\it insensitivity} to initial conditions. 
In contrast, the expelling side is very strongly divergent, with a
{\it sensitivity} to initial conditions stronger than
exponential. 
Thus, the addition of an external reinjection mechanism connecting the right
hand side with the left hand side allows an intermittent dynamics in
which chaotic and regular patterns alternate. 
Recently, a remarkable connection between 
this mechanism and the dynamics of a cluster of
spins at a thermal critical point has been established
\cite{contoyiannis}. This opened the possibility for the description
of such temporary critical fluctuations 
in terms of Tsallis' functional forms \cite{robledo_1p5}. 

\begin{figure}\begin{center}
\includegraphics[width=0.60\columnwidth]{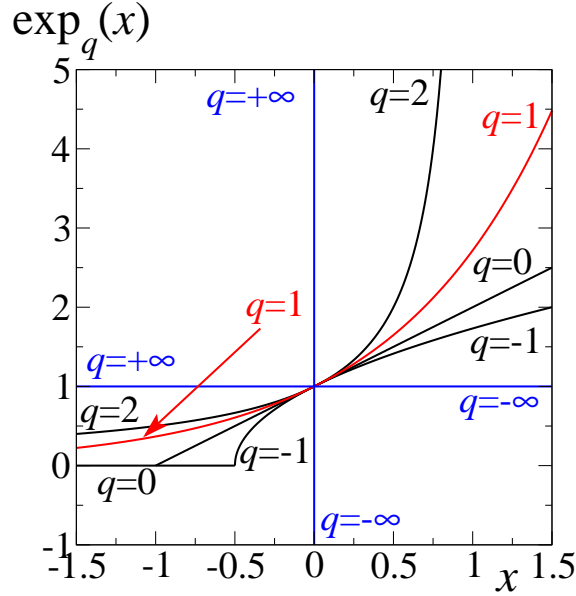}
\caption{
$q$-exponential function for positive and negative arguments and
  different values of the parameter $q\in\mathbb R$.
}
\label{fig_q_exp}
\end{center}
\end{figure}

In correspondence with the tangent bifurcation ($z=2$), Hu and Rudnick
\cite{hu} discovered that the RG iteration equation 
\begin{equation}
{f^{*}(f^{*}(x))=\tilde\alpha
    ^{-1}f^{*}(\tilde\alpha x)},{\qquad\tilde\alpha=2^{1/(z-1)}}
\end{equation}
possesses a solution, 
$f^*(x)=x[1-(z-1)u\;{\rm sgn}(x)|x|^{z-1}]^{-1/(z-1)}$,
that coincides with the expansion (\ref{eq_manneville}) up to order
$o(|x|^z)$. Remarkably, this is one of the very few RG solution
exactly known \cite{beck}. The original solution \cite{hu} can be
extended also to the pitchfork bifurcations by simply including a
change of sign \cite{baldovin_1}: 
\begin{equation} 
f^*(x)=x[1-(z-1)\;u\;(-1)^z\;{\rm sgn}(x^z)\;{\rm sgn}(x)|x|^{z-1}]^{-1/(z-1)}.
\end{equation}
Close to the bifurcation point, by using the property:
\begin{equation}
f^{*^{(m)}}(x)=\frac{1}{m^{1/(z-1)}}f^*(m^{1/(z-1)}x),\qquad m=1,2\ldots,
\end{equation}
we can recast the $m$-times iterate of the $n$-composed map
$f_{\mu,\zeta}^{(n)}$ as 
\begin{eqnarray}
x_t&\equiv& \left[f^{(n)}\right]^{(m)}(x_0)=\nonumber\\
  &=&x_0[1-(z-1)a(-1)^z{\rm sgn}(x_0^z){\rm sgn}(x_0)|x_0|^{z-1}t]^{-1/(z-1)}, 
\label{eq_ncomposed}
\end{eqnarray}
where $t\equiv m n$ and $a\equiv u/n$.
Notice that while the $n$-composed map describes a single orbit of
this form in the shifted coordinate $x_0\ll1$, for the original map
$f_{\mu,\zeta}$ there are $n$ specific time subsequences, 
each corresponding to one of these orbits originated in one of the $n$
bifurcation points. 
Eq. (\ref{eq_ncomposed}) satisfies the identity 
\begin{equation}
\frac{d x_t}{d x_0}=\left(\frac{x_t}{x_0}\right)^{z}.
\label{eq_diff}
\end{equation}
Due to this property the sensitivity to
initial conditions 
$\xi(x_0,t)\equiv\lim_{\Delta x_0\to0}\frac{\Delta x_t}{\Delta x_0}$
assumes the form 
\begin{equation}
\xi(x_0,t)=[1+(1-q)\lambda_q(x_0)t]^{\frac{1}{1-q}}\equiv
\exp_q(\lambda_q t),\qquad q\in\mathbb R.
\label{eq_qexp}
\end{equation}
The parameter $q$ can be
read from $z$. For all pitchfork (tangent) bifurcations of all order $n$
we have $q=5/3$ ($q=3/2$). 
The functional form $\exp_q$ is a one-parameter deformation of the
ordinary exponential function called 
Tsallis' $q$-exponential \cite{tsallis}.  
In the limit $q\to1$ the ordinary exponential is recovered.
The inverse of the $q$-exponential, named $q$-logarithm, is given by
$\ln_q(x)\equiv(x^{1-q}-1)/(1-q)$.
The coefficient $\lambda_q$ replaces the ordinary Lyapunov coefficient
and depends on the shifted coordinate of the initial point:
\begin{equation}
\lambda_q(x_0)=z a\;(-1)^z\;{\rm sgn}(x_0^z)\;{\rm sgn}(x_0)|x_0|^{z-1}.
\end{equation}
As for the ordinary Lyapunov coefficient, the sign of $\lambda_q$
discriminates if the dynamics is regular ($\lambda_q<0$, {\it insensitivity}
to initial conditions, meaning that two different orbits tend to
converge in time) or chaotic ($\lambda_q>0$, {\it sensitivity}
to initial conditions, i.e., two different orbits tend to
separate in time). 
In addition, the value of $q$ determines whether the dependence with
time is stronger or weaker than exponential. Fig. \ref{fig_q_exp}
reports the behavior of the $q$-exponential for different $q$'s and
sign of $\lambda_q$. 

In the next subsection we will see that a similar scenario characterizes the 
onset of chaos. Namely, the ordinary exponential form for the
sensitivity to initial conditions, which is typically independent of
the initial condition $x_0$, 
is replaced by an infinite family of Tsallis $q$-exponential
functions, each associated to a specific time subsequence, and where
the generalized Lyapunov coefficients $\lambda_q$ become
dependent on $x_0$. 
Such intricate dynamical picture is then common to all pitchfork and tangent
bifurcations, and to the periodic and quasi-periodic onset of chaos. 

\begin{figure}\begin{center}
\includegraphics[width=0.60\columnwidth]{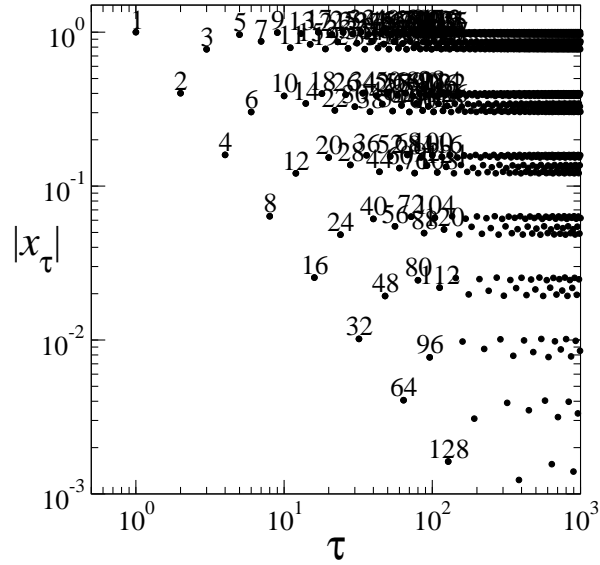}
\caption{
Time subsequences generated at the onset of chaos from
$x_{\tau=0}=0$. Numbers label the iteration time $\tau$.
Each subsequence lies along a different diagonal
line. 
}
\label{fig_subseq}
\end{center}
\end{figure}

\subsection{Onset of chaos}
\label{sub_onset}
For the sake of simplicity we address first the logistic map
$\zeta=2$. We also change slightly our notations indicating for a
moment with $\tau=0,1,\ldots$ the iteration time. Eq. (\ref{eq_unimodal})
reads then
\begin{equation}
x_{\tau+1} = 1-\mu x_\tau^2.
\end{equation}
The chaos threshold is located at $\mu_\infty(2)=1.40115\ldots$. 
Let us start by choosing the initial condition $x_0=0$.
The Feigenbaum attractor can then be characterized \cite{baldovin_2,baldovin_3} by a
series of monotonic subsequences 
\begin{equation}
\tau_k=(2k+1)2^{n-k},\quad k=0,1,\ldots,\quad n\geq k.
\end{equation}
The log-log plot of Fig. \ref{fig_subseq} shows that for each fixed
$k$, $n\geq k$ defines a subsequence such that the absolute values of
the iterates  $|x_{\tau_k}|$ lie along a diagonal line.

\begin{figure}\begin{center}
\includegraphics[width=0.60\columnwidth]{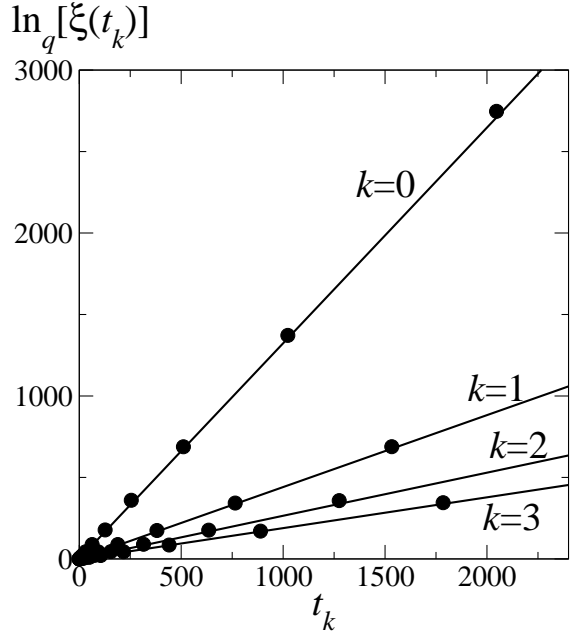}
\caption{
Sensitivity to initial condition at the onset of chaos of the logistic
map. For different subsequences, 
the $q$-logarithm/linear plot with $q=1-\frac{\ln 2}{\ln\alpha}$
exhibits straight lines whose slope corresponds to the generalized
Lyapunov coefficient $\lambda_q$. 
}
\label{fig_sen_onset}
\end{center}
\end{figure}

The Feigenbaum-Coullet-Tresser RG transformation
defines the universality class of the onset of chaos for the
logistic map. Such transformation is given by \cite{schuster,beck}  
\begin{equation}
g(x)=\alpha^{n}g^{(2^{n})}(x/\alpha ^{n}),\quad\alpha=2.50290\ldots,
\end{equation}
where $\alpha$ is one of the Feigenbaum's universal constant and
$g$ is the universal Feigenbaum's function (not known in closed form).  
This RG transformation allows to calculate explicitly the iterates for
each subsequence $\tau_k$. In fact, one obtains
\cite{baldovin_3} the expansion
\begin{equation}
x_{\tau_k}\equiv|g^{(\tau_k)}(x_0)|\simeq
  \frac{g^{(2k+1)}(0)}{\alpha^{(n-k)}}+
  \frac{g^{(2k+1)\prime\prime}(0)}{2\alpha^{(k-n)}}x_0^2,
\label{eq_onset_expansion}
\end{equation}
which is valid for initial conditions $|x_0|<\alpha^{-(n-k)}$.
We now shift to zero the initial time of each subsequence by
redefining the time as 
\begin{equation}
t_k\equiv\tau_k-2k-1=(2k+1)2^{n-k}-2k-1,\qquad n\geq k.
\end{equation}
Implementing this time shift in Eq. (\ref{eq_onset_expansion}) and
using the identity $\alpha^{(n-k)}=[1+t_k/(2k+1)]^{\ln\alpha/\ln2}$,
the sensitivity to initial conditions
for each time subsequence turns out to be  \cite{baldovin_3} 
\begin{equation}
\xi_{t_k}=\exp_q\left[\lambda_{q}^{(k)}t_k\right],
\quad q=1-\frac{\ln 2}{\ln\alpha},
\quad\lambda_{q}^{(k)}=\frac{\ln\alpha}{(2k+1)\ln2}
\label{eq_sen_onset}
\end{equation}
(see Fig. \ref{fig_sen_onset}). 
The fact that the generalized Lyapunov coefficient $\lambda_{q}^{(k)}$
varies with the subsequence $k$ can be seen as a dependence of this
coefficient on the initial condition. In fact, for each subsequence, the
initial position at zero time $t_k$, $x_{\tau=2k+1}$, changes with $k$.

Along these lines, it is also possible to show \cite{baldovin_3} that within each
subsequence, Tsallis' $q$-entropy grows linearly with the time $t_k$
and the slope of the linear growth is equal to\footnote{
The reader may be interested by the fact that in a different context,
namely for a conservative bidimensional map with vanishing Lyapunov
coefficients, similar numerical evidence has been found, although for
the somewhat trivial (linear) case $q=0$ \cite{casati}.}
$\lambda_{q}^{(k)}$. This, parallels the Pesin identity that relates
the (sum of the) positive Lyapunov coefficient(s) to the Kolmogorov-Sinai
entropy rate \cite{beck}, and what happens for the ordinary entropy
growth ($q=1$) when strong chaos is present.

Although Eq. (\ref{eq_sen_onset}) gives the sensitivity to initial
conditions for all the subsequences $k$, it is not yet completely
general since such subsequences are those generated starting at
$x_{\tau=0}=0$.  
At this point, we have to say what happens if we choose
another point of the attractor as the starting point.
The answer is given by considering
the Feigenbaum's $\sigma$
function. This function is defined by the scaling properties of the
diameters of the bifurcation forks $d_{r,s}$:
\begin{equation}
\sigma_r(s)\equiv\frac{d_{r+1,s}}{d_{r,s}}.
\end{equation}
In the limit $r\to\infty$, $\sigma_r$ has an infinite number of
jumps, each associated to a different attractor point which is
here identified by the variable $s$. 
The sensitivity to initial conditions for the subsequence $k$
generated by the initial point $x_{\tau=0}$ defined by $s$ can be
deduced from $\sigma$ through \cite{mayoral_1}
\begin{equation}
\xi_t(x_0)=\lim_{n\to\infty}\left|\frac{d_{n-k+1,s+t}}{d_{n-k,s}}\right| 
\sim\left|\frac{\sigma_{n-k}(s-1)}{\sigma_{n-k}(s)}\right|^{n-k}=A^{n-k},
\end{equation}
where $A$ is a number that depends on $s$ (and therefore on
$x_{\tau=0}$). 
The quantity $A$ has been identified as the ratio of the
values of $\sigma$ at each of its discontinuities \cite{mayoral_1}.
Hence, again the use of the identity
${ A^n=[1+\frac{t_k}{2k+1}]^\frac{\;\ln A}{\ln2}}$ lead to the more
general result
\begin{equation}
q=1-\frac{\ln2}{\ln A},\qquad\lambda_q=\frac{\ln A}{(2k+1)\ln2}.
\end{equation}
We see then that the structure valid for $x_{\tau=0}=0$ is maintained,
but changing the starting points implies a change of the power law
exponent $q$. 

Finally, all previous results for the onset of chaos can be further
generalized to general nonlinearity $\zeta$ in unimodal maps
\cite{mayoral_1}, in which case the families of $q$-exponential functions
characterizing the sensitivity to initial conditions are defined by 
\begin{equation}
q=1-\frac{\ln2}{(\zeta-1)\ln A},\qquad
\lambda_q=\frac{(\zeta-1)\ln A}{(2k+1)\ln2},
\end{equation}
where $A$ now also depends on $\zeta$.

Summarizing, we have seen that the sensitivity to initial conditions at
the chaos threshold can be understood in terms of a complex
hierarchical structure that can be exactly deduced from the
Feigenbaum-Coullet-Tresser RG approach. For a given point of the
attractor, this structure corresponds to an infinite family of
$q$-exponentials, each with the same $q$ but with a different
generalized Lyapunov coefficient $\lambda_q^{(k)}$. Spanning the
initial point along the attractor corresponds to an infinite number of
such families, each of the families being now
characterized by a different value of $q$ which is given in terms of
the discontinuities of Feigenbaum $\sigma$ function. Since the
stronger of such discontinuities happens for $s=0$ ($x_{\tau=0}=0$), 
the value of $q$ corresponding to this discontinuity is in some sense
the most important one. Such a value, for the logistic map, is
$q=1-\ln2/\ln\alpha=0.2445\ldots$. 

Another fundamental result \cite{mayoral_1} for the period doubling
onset of chaos (and actually also for the quasi periodic route to
chaos \cite{saldana_1}) is
that the Tsallis' $q$-exponential structure for the dynamics can be
exactly linked with the formalism of Mori's transitions
\cite{mori_2}, which we briefly sketch below. 
Mori's formalism, when applied to the transition to chaos, 
starts by defining a different generalization of
the Lyapunov coefficient, called {\it generalized finite-time Lyapunov 
coefficient} 
\begin{equation}
\lambda(x_0,t)=\frac{1}{\ln t}\sum_{i=0}^{t-1}
\ln\left|\frac{d f_{\mu_\infty}(x_i)}{d x_i}\right|.
\end{equation}
The probability for getting the value $\lambda$ at the finite-time $t$
is postulated in the form
\begin{equation}
p(\lambda,t)\equiv t^{-{\psi(\lambda)}}p(0,t).
\end{equation}
This allows the introduction of a dynamic partition function
$Z(\mathbf{q},t)$ and of the associated free energy 
${\phi(\mathbf q)}$ 
\begin{equation}
Z(\mathbf{q},t)\equiv \int d\lambda p(\lambda,t)t^{-({\mathbf q}-1)\lambda},
\quad{\phi(\mathbf q)}\equiv-\lim_{t\to\infty}\frac{\ln Z(\mathbf q,t)}{\ln t},
\end{equation}
where $\mathbf q\in\mathbb R$ is Mori's $\mathbf q$-parameter. 
The ``coarse grained'' function of generalized finite-time Lyapunov
coefficients is
\begin{equation}
\lambda(\mathbf q)\equiv \frac{d\phi(\mathbf q)}{d\mathbf q},
\end{equation}
and ${\psi(\lambda)}$ is the Legendre transform of  
${\phi(\mathbf q)}$. 
In analogy with ordinary thermal phase transitions, Mori and
colleagues detected a single first order phase transition for the
onset of chaos \cite{mori_2}. 

The important identification \cite{robledo_3} 
\begin{equation}
\lambda\equiv\frac{1}{\ln n}\ln A^{n-k}=\frac{1}{n}\ln_q A^{n-k},
\end{equation}
together with the previous results, leads to the conclusion
\cite{mayoral_1}  that the
dynamics at the onset of chaos is actually characterized by an 
{\it infinite family} of Mori's $\mathbf q$-phase transitions each
occurring at one of the allowed values of the Tsallis' $q$-index in
the sensitivity: 
\begin{equation}
\mathbf q=q.
\end{equation}

\subsection{Quasiperiodic onset of chaos}
The prototype for the quasiperiodic route to chaos is the circle map
(see, e.g., \cite{ott})
\begin{equation}
\theta_{t+1}=f_{\Omega,K}(\theta_t)=\theta_t+\Omega-\frac{K}{2\pi}\sin(2\pi\theta_t)\qquad
  (\theta\;{\rm mod}\;1),
\end{equation}
which depends on the real parameters $\Omega$ (bare winding number) and
$K$ (amount of nonlinearity). In order to describe the dynamical
behavior it is important to consider the influence of the parameter
$\Omega$ through the {\it dressed winding number}
\begin{equation}
w\equiv\lim_{t\to\infty}\frac{\theta_t-\theta_0}{t},
\end{equation} 
that gives the average increment of $\theta_t$ per iteration. 
The orbits for $K<1$ are periodic (locked motion) if $w$ is rational
and quasiperiodic (unlocked motion) if $w$ is irrational. 
In fact, $w(\Omega)$ describes the so-called ``devil staircase''
making a step at each rational value of $\Omega$ \cite{ott}.
$K=1$ defines the {\it critical circle map} for which the periodic
motion covers the whole $\Omega$ interval apart from a multifractal
set of values for which the motion is quasiperiodic.
For $K>1$ regions of periodic motion (Arnold tongues) overlap in
chaotic bands. 
The chaos threshold can be studied by fixing $K=1$ and by approximating
a target irrational value of $w$ through a sequence or rational
values. 
Perhaps the most interesting case is the one in which the target value is
the reciprocal of the golden mean $w_{gm}\equiv(\sqrt{5}-1)/2$ and the
sequence of rationals is given by Fibonacci numbers
$w_n\equiv F_{n-1}/F_n$ (with $F_{n+1}=F_n+F_{n-1}$) \cite{ott}. 
The universal properties of the onset of chaos are then described by
the fixed-point map $g_{gm}(\theta)$ of the following RG transformation
\begin{equation}
g_{gm}(\theta)=\alpha_{gm}\;
g_{gm}\left(\alpha_{gm}g_{gm}\left(\frac{\theta}{\alpha_{gm}^2}\right)\right),
\end{equation}
where $\alpha_{gm}=-1.288575\ldots$ is a universal constant. 
The same techniques implemented for the pitchfork bifurcation onset of
chaos lead to the previous universal picture, although
this transition to chaos is much more involved and reach in detail
\cite{saldana_1}.  
Namely, the sensitivity
to initial condition at the chaos threshold is given by an infinite
number of Tsallis' $q$-exponential families, each $q$-exponential
being associated to a specific time subsequence\footnote{
More precisely, the time subsequences are in this case identified by two
different numbers $k=1,2,\ldots$ and
$l=0,1,\ldots,k-1$. Eq. (\ref{eq_sen_circle}) is valid if $l=0$ (see
\cite{saldana_1} for 
details). 
}:
\begin{equation}
\xi_{t_k}=\exp_q\left[\lambda_{q}^{(k)}t_k\right],
\quad q=1+\frac{\ln w_{gm}}{2\ln |A|},
\quad\lambda_{q}^{(k)}=2\frac{\ln |A|}{k\ln w_{gm}},
\label{eq_sen_circle}
\end{equation}
where $A$ depends on the initial condition that generates the time
subsequence $k$ ($A=\alpha_{gm}$ if the initial condition is 
at $\theta_0=0$). 
Correspondingly, the dynamics is characterized by an infinite
family of Mori's $\mathbf q$-transitions occurring at 
$\mathbf q=q$ \cite{saldana_1}.

\subsection{Tsallis' $q$-exponential sensitivity to initial conditions}
We have seen that if one is interested in extracting significant trends
for the dynamics at critical attractors where the ordinary Lyapunov
coefficient vanishes, some precise time subsequences that depend on
the initial condition must be selected. Within these subsequences,
Tsallis' $q$-exponentials and $q$-generalized Lyapunov coefficients
$\lambda_q$ appear to be convenient tools for extracting these
trends. Table \ref{table_q_exp} summarizes the different behaviors obtained
from various combinations of $q$ and $\lambda_q$. 
The only case not yet observed is $q<1$ and
$\lambda_q<0$, which would amount to super-strong (stronger than
exponential) insensitivity to
initial conditions. In private conversations with A. Robledo, we
conjectured that this could be the case of the super-stable cycles of
unimodal maps. Efforts to clarify this point are currently in course
\cite{saldana_2}.  

\begin{table}[h]
\begin{center}
\begin{tabular}[c]{|c||c|c|c|}
\hline
& $q<1$ & $q=1$ & $q>1$ \\
\hline\hline
$\lambda_q<0$ & {\it super-strong}&
{\it strong} & {\it weak} \\
 & {\it insensitivity} & 
{\it insensitivity}  & {\it insensitivity} \\
 & super-stable cycles?& $\mu<\mu_\infty$ & pitchfork, tangent lhs \\
\hline
$\lambda_q>0$ & {\it weak} &
{\it strong} & {\it super-strong} \\
 & {\it sensitivity} & 
{\it sensitivity} & {\it sensitivity} \\
 & $\mu=\mu_\infty$ & typically for $\mu>\mu_\infty$ & tangent rhs \\
\hline
\end{tabular}
\caption{
Sensitivity to initial conditions and Tsallis' $q$-exponentials.
}
\label{table_q_exp}
\end{center}
\end{table}

\section{Logistic map with additive noise and glassy dynamics}
\label{glassy}
The theoretical understanding of the
dynamical and thermodynamical features of supercooled liquids
is nowadays a very important research topic in condensed
matter physics and in statistical
mechanics. Theoretical efforts are motivated by the search for 
general common features of non-equilibrium systems and applications span from
new material science to protein folding.   
In a glass former there are two different dynamical
scenarios, separated by the glass
transition temperature $T_g$. 
Above $T_g$
the system equilibrates while
below $T_g$ it is out-of-equilibrium.
Phenomenology of glassy systems includes \cite{debenedetti}: 
\begin{itemize}
\item Time translation invariant relaxation processes. Above, but
  close to, the glass
  transition two-time correlations do not depend on the first
  (waiting) time and display power law decays ($\alpha$-decays). Some
  glass formers also exhibit a further initial power law decay called
  $\beta$-decay. 
\item Loss of time translation invariance (aging) at and below the glass
transition.
\item Intriguing connection between kinetics and
thermodynamics -- Adam-Gibbs formula:  
\begin{equation}
t_x=A\exp\left(\frac{B}{TS_c}\right),
\end{equation}
where $t_x$
is a relaxation time 
(equivalently the
viscosity), $A$ and $B$ are constants, $T$ is temperature,
and $S_c$
configurational entropy
\item Critical slowing down: transition from
normal- to sub-diffusion
to localization of molecular motion.
\end{itemize}

In order to elaborate the counterpart of such behaviors in
low-dimensional maps, 
let us start by noticing that in the Langevin theory the effect of
collisions of the diffusive particle 
with molecules in the fluid  is represented by an additive noise term.
In the same spirit, nonlinear low-dimensional maps with
external noise can be used to model systems with many
degrees of freedom.
The discrete form for a Langevin equation is
\begin{equation}
x_{\tau+1}=x_\tau+h_\mu(x_\tau)+\sigma\Gamma_\tau\qquad \tau=0,1,\ldots,
\label{eq_langevin}
\end{equation}
where $\Gamma_\tau$ is a Gaussian white noise
($\langle\Gamma_{\tau}\Gamma_{\tau^{'}}\rangle=\delta_{\tau,\tau^{'}}$)
and $\sigma$
measures the noise intensity.
Taking $h_\mu(x)=1-x-\mu x^2$,
Eq. (\ref{eq_langevin}) becomes the iteration equation of the Logistic
map with additive noise
\be
x_{\tau+1} =  1-\mu
x_\tau^2+{\sigma \Gamma_\tau},
\quad x\in[-1,1],\quad\mu\in[0,2].
\ee
For small noise amplitudes the onset of chaos $\mu_\infty(\sigma)$
still separates an
``equilibrium'' ergodic phase characterized by a chaotic dynamics with
positive Lyapunov coefficient from an ``out of equilibrium''
non-ergodic one where
the Lyapunov coefficient is negative \cite{schuster}.

\begin{figure}
\begin{center}
\includegraphics[width=0.48\columnwidth]{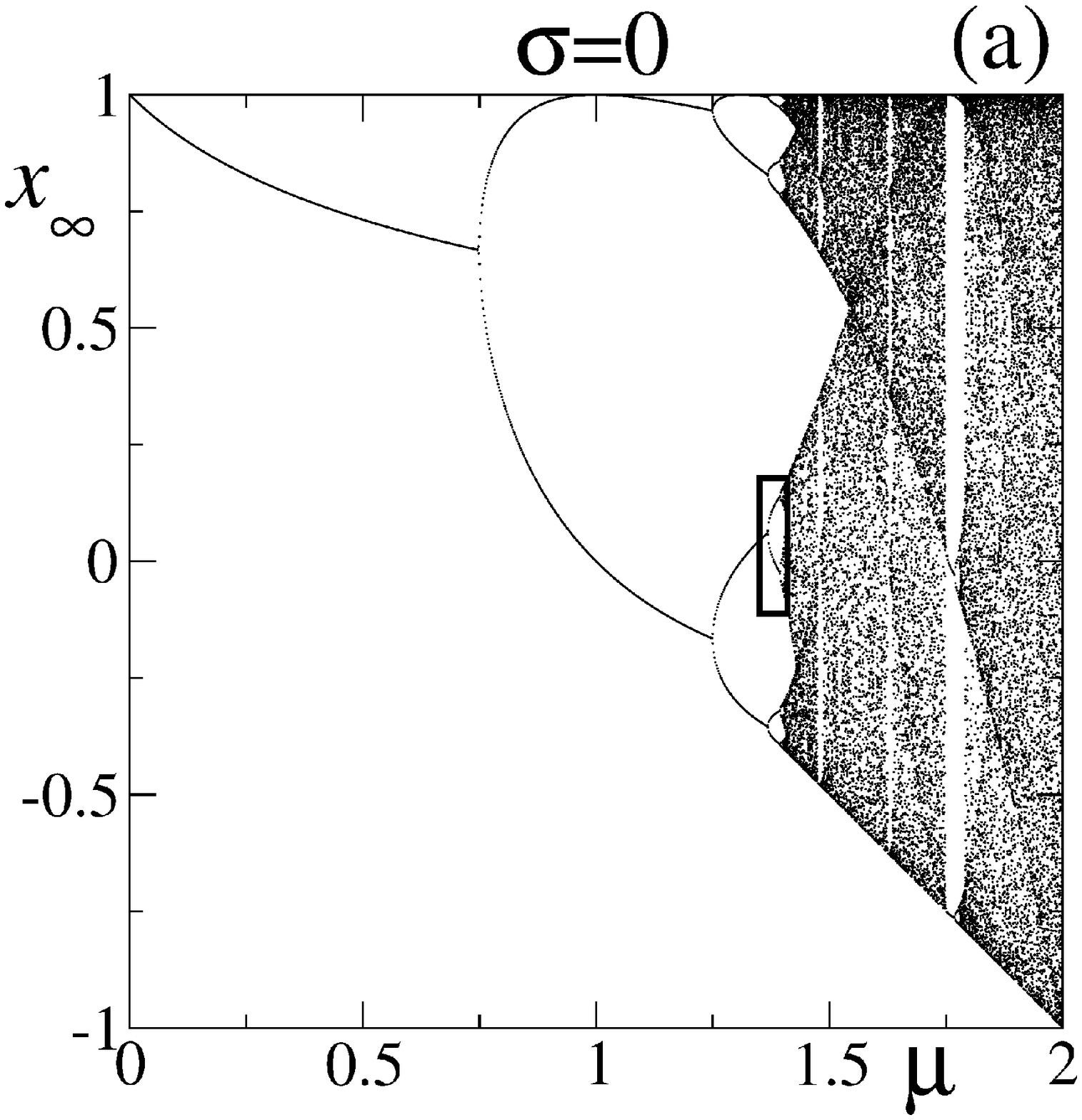}
\includegraphics[width=0.48\columnwidth]{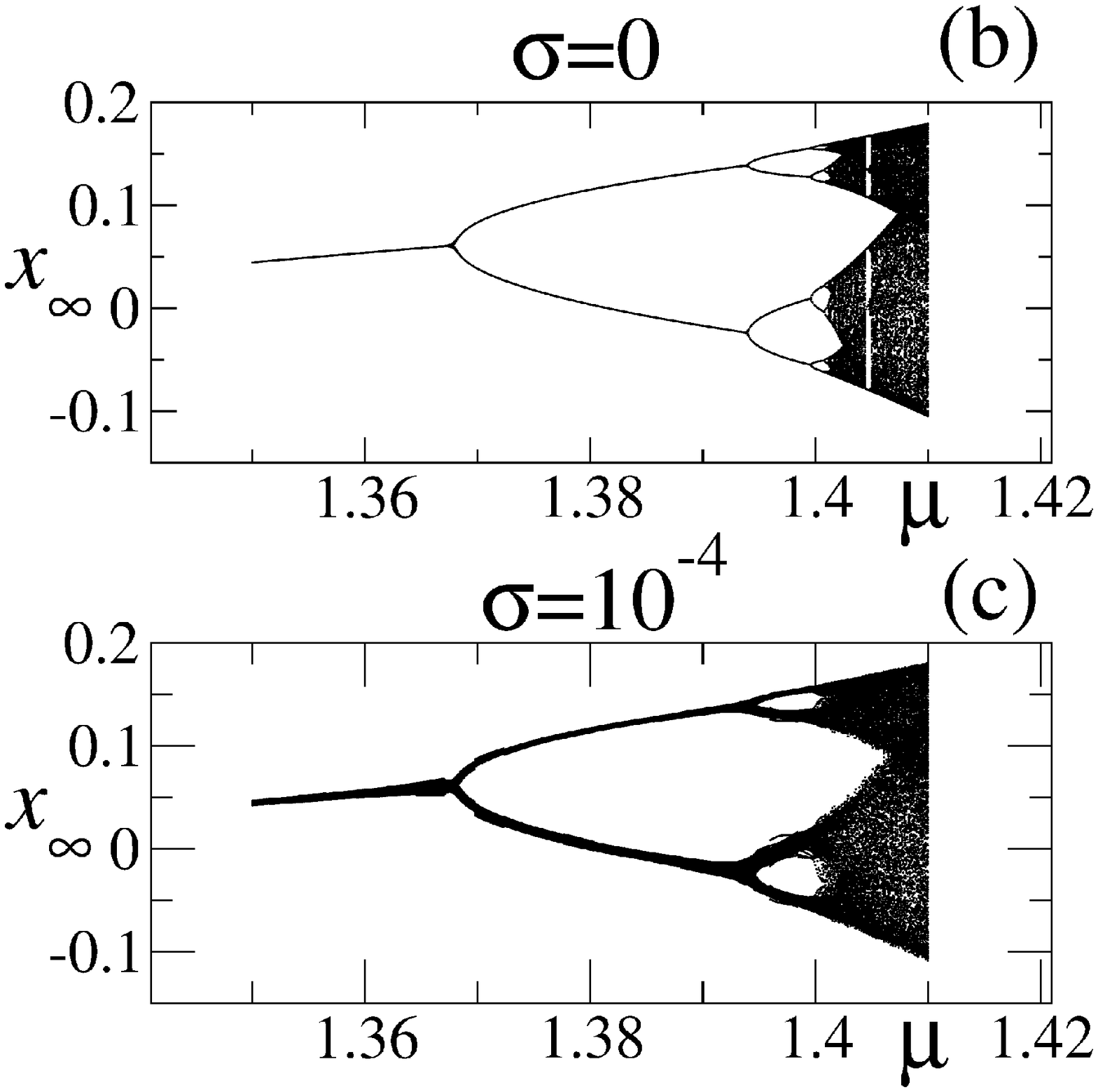}
\caption{
Bifurcation gap phenomenon. (b) and (c) are magnification of the region
enclosed by the box in (a). The infinite number of bands (points) at the
noiseless chaos threshold in (b) become a finite number of
chaotic bands in presence of the noise in (c).
}
\label{fig_bifurcation_gap}
\end{center}
\end{figure}

We will be studying the chaos threshold from $\mu>\mu_\infty$. In
our analogy with the glass former this means that we will be looking
at the glass formation from the side of the liquid. 
In the absence of noise, the attractor, besides being the accumulation
point of the bifurcations cascade, can also be viewed as the
accumulation point of the band splittings occurring above the
onset of chaos. For $\mu=2$, the chaotic attractor is in fact characterized by
a single band that spans the interval $[-1,1]$ (see
Fig. \ref{fig_bifurcation_gap}a). As $\mu$ is reduced towards
$\mu_{\infty}$, this unique band reduces its width and then splits in
two disjoint bands. This band splitting continues up to $\mu_\infty$,
where the attractor can be viewed as $2^\infty$ disjoint
bands. 
The same occurs if $\sigma\neq0$, the only difference being that 
the band splittings end (and the
bifurcations end) at a finite value
$2^{N(\sigma)}$ (Fig. \ref{fig_bifurcation_gap}b). 
This phenomenon is called {\it bifurcation gap}. 

\begin{figure}\begin{center}
\includegraphics[width=0.60\columnwidth]{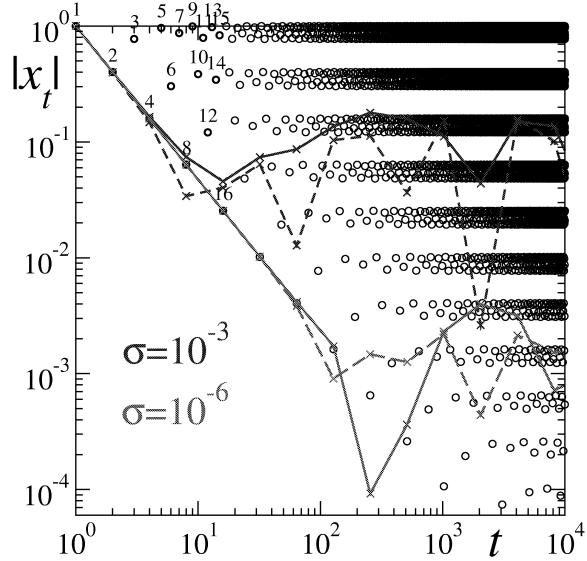}
\caption{
Crossover dynamics observed for the main subsequence $k=0$ for two
different values of the noise intensity. For $t<t_x(\sigma)$ the
iterates closely follow the noiseless attractor. For $t>t_x(\sigma)$
the iterates become more chaotic and they jump among different bands
of the noiseless attractor.  
}
\label{fig_crossover}
\end{center}
\end{figure}

If the noise term is small enough, the RG perturbative approach can be
applied in order to calculate the position of the iterates around
$x_0=0$ \cite{robledo_2}. The result is  
\be
x_{\tau_k}=
  \tau_k^{-\frac{1}{1-q}}\left
  |g\left(\tau_k^{\frac{1}{1-q}}x_0\right)
  { +\sigma\Gamma_{\tau_k} \tau_k^{\frac{1}{1-r}}}+
  G_\Lambda\left(\tau_k^{\frac{1}{1-q}}x_0\right)
  \right|,
\ee
where $|x_0|\ll1$, $G_\Lambda(x)$ 
is the first order perturbation
eigenfunction, $r\simeq0.6332$
is determined from the
noisy power spectrum of the map \cite{schuster}, 
and we are using a notation consistent
with the one in Section \ref{sub_onset}. For the leading subsequence $k=0$, by
introducing the previous time shift $t\equiv t_0=\tau_0-1$, the iterate positions 
can be viewed in terms of $q$-exponentials:
\be
x_{t}=\exp_{2-q}\left(-\lambda_{q}t\right)
\left[1{ +\sigma\Gamma_t\exp_{r}\left(-\lambda_{r}t\right)}\right].
\ee
This implies the presence of a ``crossover'' or ``relaxation'' time in
the noisy dynamical behavior:
\be
t_x=\sigma^{r-1}.
\ee
For $t<t_x=2^{N(\sigma)}-1$ the dynamics is restricted to non-overlapping bands
attractor, similarly to the noiseless case; 
for $t>t_x$ the dynamics becomes more and more chaotic
as a consequence of the band merging (Fig. \ref{fig_crossover}).

This crossover furnishes the key for
identifying the following marks of glassy-like behavior for the noisy
onset of chaos (see \cite{robledo_2,baldovin_4} for details). 
\begin{itemize}
\item {\it Analogue of the Adam-Gibbs formula.}
The entropy associated to the distribution of the
iterate positions within the $2^N$ bands has the form
\be
S_c=2^N\sigma s,
\ee
where $\sigma s\equiv-\sigma\int d\Gamma\;p(\Gamma)\ln p(\Gamma)$ is
the entropy associated to a single band.
Since $2^N=1+t_x$ and $t_x=\sigma^{(r-1)}$ we get \cite{robledo_2}:
\be
t_x=\left(\frac{s}{S_c}\right)^{\frac{1-r}{r}}.
\ee
Notice that in contrast with the Adam-Gibbs exponential law, in this
case $t_x$ has a power law behavior in $1/S_c$.

\item {\it Time translation invariance and $\alpha$-relaxation.}
Whereas a study of the relaxation processes of iterates starting with
initial conditions $x_0$ outside the attractor could furnish some
analogue to the $\beta$ initial fast relaxation observed for some
glass formers, the $\alpha$ subsequent relaxation is mimicked
by the bifurcation-gap crossover for $t>t_x$.
This becomes apparent if one studies the behavior of ensemble-averaged
correlations $c_{e}(t_{1},t_{2})$ \cite{baldovin_4}. 
There is a ``thermalization'' process occurring for
small values of $t_1$. If $t_1$ is large enough inside the chosen
subsequence $k$, the correlation becomes
time translation invariant and $t_x$ characterizes the decay of
correlations.

\item {\it Aging.}
At the onset of chaos when $\sigma=0$ trajectories are
non-ergodic and retain memory of the initial data. 
This property is equivalent to a loss of time translation invariance
and to a ``built-in'' aging.
Since ensemble and time averages are not equivalent, a
time-averaged definition of correlations 
allows the exhibition of such a property \cite{baldovin_4}.

\item {\it Subdiffusion and arrest.}
The sharp slowing down of dynamics in supercooled liquids implies a
crossover from normal diffusion to sub-diffusion and finally to
arrest. 
This deceleration is caused by the confinement of any molecule by a
``cage'' formed by its neighbors.
Such a ``cage'' can be reproduced by a periodic map obtained via
repetition of a single cell map in such a way that diffusion is due
only to the noise term. 
The periodic map is given by 
\be
x_{t+1}=F(x_{t}),\quad F(l+x)=l+F(x),\quad l=...-1,0,1,\ldots, 
\label{eq_subdiffusion}
\ee
where 
\begin{equation}
F(x)=\left\{ 
\begin{array}{c}
-\left\vert 1-\mu _{c}x^{2}\right\vert +\sigma \xi ,\;-1\leq x<0, \\ 
\left\vert 1-\mu _{c}x^{2}\right\vert +\sigma \xi ,\;0\leq x<1.
\end{array}
\right.  \label{cellmap1}
\end{equation}
The diffusion process displayed by the map in
Eq. (\ref{eq_subdiffusion}) reproduces the crossover
normal/sub-diffusion/arrest as $\sigma$ tends to zero \cite{baldovin_4}.  
\end{itemize}

\section{Final remarks}
Several nonlinear dissipative systems, like those mentioned in this
review, exhibit both regular and chaotic behavior depending on some
external parameter. 
The transition from regularity to chaoticity 
implies a dramatic alteration in the dynamical behavior,
the situation changing from an exponentially fast {\it insensitivity} to 
initial conditions to an exponentially fast {\it sensitivity} to initial
conditions.
There are only few possible mechanisms through
which such transitions occur. 
These are the well known (universal)
routes to chaos. 
Thus, when the external parameters are tuned {\it at} the transition
point, the ordinary Lyapunov coefficients vanish and they do not allow
to extract significant trends characterizing the critical dynamics. 
Under these conditions, intricate oscillations typically appear. Since
these oscillations are reminiscent of the regular behavior, the
possibility arises to characterize the dynamics exactly if one accepts to
describe it via a set of specific temporal subsequences. 

Here we have given an account of the extension of the static RG
approaches, discovered in the 80's, to the dynamics at the critical
attractors of period doubling, intermittency and quasiperiodicity
routes to chaos
\cite{robledo_1,baldovin_1,baldovin_2,baldovin_3,mayoral_1,saldana_1}.
The outcome of this extension, obtained
basically from an {\it a priori} approach, are significant new results
that enlarge considerably the previous knowledge. For example, it is
now clear that all the properties found for the sensitivity to initial
conditions are associated to the discontinuities of Feigenbaum's
universal $\sigma$ scaling function for the bifurcation cascade onset of
chaos \cite{mayoral_1} (with an analogous extension for the
quasiperiodic critical attractor \cite{saldana_1}). 
In the language of Tsallis' functional forms these
properties can be read as Tsallis' generalization of the Lyapunov
coefficient that exhibits a new spectrum of behaviors ranging from
weaker to stronger than exponential laws
\cite{baldovin_1,baldovin_2,baldovin_3}. Also, precise contact
with the formalism of Mori's dynamical transitions at the transition
point and the fixed $q$-parameter in Tsallis' formalism
has been established \cite{mayoral_1,saldana_1}. In view of
this relation on can conclude that the dynamics for the period
doubling and the quasiperiodic critical attractors is in fact
characterized by an infinite number of Mori's dynamical transitions
\cite{mayoral_1,saldana_1}, in place of the single one known
previously. In relation to a recent debate in the literature
\cite{grassberger_1}, let us then conclude that the 
novel exact knowledge on the critical dynamics obtained in 
\cite{robledo_1,baldovin_1,baldovin_2,baldovin_3,mayoral_1,saldana_1}
puts into a different fresh context the fundamental scaling laws
discovered in the 80's for these nonlinear dynamical systems.

As an example of how to implement the detailed knowledge of the
dynamics at the period doubling onset of chaos, we have reported
\cite{robledo_2,baldovin_4} that
specific parallels with the behavior of a glass former close to the
glass transition can be established if one adds a noise term to the
logistic map equation. 
Of course through this very simple
one-dimensional system we can aim only at qualitative analogies
with respect to the behavior of a supercooled liquid close to the glass
transition, nevertheless these similarities have the appeal of
allowing a precise numerical and analytical inspection. 

This paper is dedicated to Alberto Robledo, with whom 
I had the great privilege and pleasure of sharing
personal and professional views during long term collaborations and
engaging discussions.

\section*{Acknowledgments}
I thank E. Orlandini, A.L. Stella and B. Marcone for useful
remarks, and M. Zamberlan for daily support.

% \appendix

% \section{}
% \label{}

\end{document}